\pdfoutput=1
\documentclass[referee]{raa}            

\usepackage{graphicx,times}             
\usepackage{natbib}
\usepackage{amssymb,amsmath}
\bibpunct{(}{)}{;}{a}{}{,}

\usepackage[pagebackref=true]{hyperref}
\begin{document}

  \title{Study of Central Intensity Ratio of Seyfert Galaxies in nearby Universe 
}

   \volnopage{Vol.0 (20xx) No.0, 000--000}      
   \setcounter{page}{1}          

   \author{K T Vinod 
      \inst{1}
   \and C Baheeja
      \inst{1}
   \and S Aswathy
      \inst{2}
   \and C D Ravikumar
      \inst{1}
   }

   \institute{Department of Physics, University of Calicut, Malappuram-673635, India; {\it vinod2085@gmail.com}\\
        \and
             Department of Physics, Providence Women's College, Calicut-673009, India\\
\vs\no
   {\small Received 20xx month day; accepted 20xx month day}}

\abstract{ We use the recently discovered simple photometric parameter Central Intensity Ratio  \citep[CIR,][]{2018MNRAS.477.2399A} determined for a sample of 57 nearby ($z<0.02$)  Seyfert galaxies to explore the central features of galaxies and their possible connection with galaxy evolution. The sample of galaxies shows strong anti-correlation between CIR and mass of their central supermassive black holes (SMBH).  The SMBH masses of ellipticals are systematically higher for a given CIR value than that for lenticulars and spirals in the sample. However, the correlation between CIR and central velocity dispersion is weak. CIR appears less influenced by the excess flux produced by the central engine in these galaxies, when compared to spectroscopic parameters like velocity dispersion and OIV flux, and proves a fast and reliable tool for estimating central SMBH mass.
}

   \authorrunning{Vinod et al. }            
   \titlerunning{Central Intensity Ratio of Seyfert galaxies }  

   \maketitle

%
%
\section{Introduction}           
\label{sect:intro}

The central supermassive black hole (SMBH) residing in massive galaxies is believed to play a key role in the evolution scenario of host galaxies. The evolution mechanism of the central engine of every galaxy is connected with the star formation process in the host galaxy. It is commonly accepted that the accretion mechanism is the prime reason for the origin and growth of active galactic nuclei (AGN) in the nuclear region of galaxies \citep{2006ApJ...637..104K,2011MNRAS.418.2043E,2011ApJ...743....2S,2012MNRAS.426..360V}. 

AGNs are hosted at the centre of elliptical galaxies or bulge dominating spheroids across all redshifts \citep{2003MNRAS.346.1055K,2009ApJ...706..810P}, whereas the morphology of local Seyfert galaxies is generally spiral \citep{1995ApJS...98..477H,2008ARA&A..46..475H}. Intense circumnuclear star formation plays a crucial role in the evolution and emission process of Seyfert galaxies, specifically, Sy2 galaxies. \citep[e.g.,][]{1985MNRAS.213..841T,1998ApJ...493..650M,1995MNRAS.272..423C,2001ApJ...546..845G,2001ApJ...558...81C}.

Seyfert galaxies are among the most studied objects in the radio-quiet (RQ) category, along with quasars \citep{1977ARA&A..15...69W,1993ApJ...414..552O,2015MNRAS.454.2918R}. The role of feedback by the central SMBH in the relationships between the mass of the SMBH and bulge properties of Seyfert galaxies is still unclear because the merger events govern the formation of bulges while Seyfert galaxies are believed to be evolving through secular evolution \citep{2006ApJS..163....1H,2013ARA&A..51..511K,2014ARA&A..52..589H}. Recent studies revealed the existence of fast outflows of ionized gas in nearby Seyfert galaxies, but their influence on star formation is still under debate \citep{2006A&A...452..869C,2007A&A...467.1037K,2012ApJ...756..180W,2014A&A...567A.125G,2015A&A...580A...1M,2016A&A...593A.118Q}. However, if the host galaxies possess such outflows, they could expel the gas from the central region and suppress the star formation \citep{ALEXANDER201293,2014A&A...567A.125G,2015ApJ...798...31A,2016MNRAS.458..816H,2016MNRAS.461.3724W}.

The masses of central supermassive black holes (SMBHs) are reported to correlate well with the stellar mass and stellar velocity dispersion of the bulges of their host galaxies \citep[see, e.g.,][]{1998AJ....115.2285M,2000ApJ...539L...9F,2000ApJ...539L..13G,2003ApJ...589L..21M,2004ApJ...604L..89H,2013ARA&A..51..511K,2013ApJ...764..184M,2015yCat..74462330S}. The bulges  and supermassive black holes seem to evolve together and regulate each other \citep{2013ApJ...765...78A}. The relations (between $\rm M_{BH}$, bulge mass, and stellar velocity dispersion) propose a strong connection between the formation of black hole mass, emergence of AGNs, and the host galaxy evolution \citep{2000ApJ...539L...9F,2009ApJ...698..198G} as well. 

Central light concentration is a vital parameter, which can be used as a tracer of the disk to bulge ratio, star formation activity, and galaxy evolution \citep{1994ApJ...432...75A,2003ApJS..147....1C}. The Central Intensity Ratio (CIR), a new photometric parameter, is well correlated with the masses of central SMBHs  of the spheroid of early-type galaxies \citep{2018MNRAS.477.2399A}. Furthermore, CIR is an efficient photometric tool to study the central and structural properties of spiral galaxies, especially barred systems, and also gives some valid information regarding nuclear star formation and AGN formalism in host galaxies \citep{2020RAA....20...15A}. In this light, we perform an optical analysis by utilizing the parameter, CIR, to study the central properties and evolution of Seyfert galaxies. 

This paper is organized as follows; Section 2 describes the properties of the sample galaxies and the data reduction techniques employed in this study, and Section 3 deals with results consisting of various correlations. Discussions and conclusions are provided in Section 4.

Throughout this paper, we have used the cosmological parameters: $ H_0=73.0\,  {\rm km \, s^{-1}}  {Mpc}^{-1}$; $\Omega_{\rm matter} = 0.27$; $\Omega_{\rm vacuum} =0.73$.


\begin{table}
\addtolength{\tabcolsep}{-3pt}
\begin{center}
\caption[]{ Table 1 lists the properties of sample galaxies. Name of the galaxy (column 1), Distance (2), Seyfert type (3), Morphology (4), CIR computed in the corresponding filter (5), uncertainty of CIR (6), SMBH mass (7) and corresponding references (8), stellar velocity dispersion adopted from Hyperleda (9), OIV flux taken from \cite{2009ApJ...698..623D} (10), estimated circumnuclear SFR (11), uncertainty of SFR (12), HST observation (13).}\label{Tab1}


\begin{tabular}{lcccccccccccc} 
\hline
\vspace{-1mm}
Galaxy & Distance & Seyfert & Morphology & CIR & $\Delta$CIR & log $M_{\rm BH}$ & ref & $\sigma$  & OIV flux  & log SFR & $\Delta$SFR & HST  \\
 &(Mpc) & type &  & &  & $( M_{\odot})$ &  &  $\rm{(km\,s^{-1})}$ &$\rm{(erg\,cm^{-2}\,s^{-1})}$& $(M_{\odot}\,yr^{-1})$ & & obs.  \\
\hline
IC2560&	40.7& S2&	(R')SB(r)b&	1.44&	0.03&	6.64&	1&	136.5& 5.43E-013&	0.67&	0.013&	WFPC2\_F814\\
IC3639&	35.3&	S2&	SB(rs)bc&	1.27&	0.02&	6.83&	2&	97.1&	3.55E-013&	1.64&	0.040&	WFPC2\_F606\\
NGC0613&	20.7&	S?&	SB(rs)bc&	0.86&	0.02&	7.60&	3&	122.1&	$-$ &	0.91&	0.007&	WFPC2\_F814\\
NGC0788&	54.1&	S2&	SA(s)0/a&	1.16&	0.03&	7.51&	2&	134.4&	1.80E-013&	1.01&	0.021&	WFPC2\_F606\\
NGC1275&	70.1&	S2&	E-cD&	1.07&	0.03&	8.58&	2&	244.6&	1.85E-013&	2.22&	0.170&	WFPC2\_F702\\
NGC1358&	53.6&	S2&	SAB(r)0/a&	1.17&	0.02&	7.88&	2&	215.1&	7.61E-014&	$-$ & $-$ &		WFPC2\_F606\\
NGC1365&	21.5&	S1.8&	SB(s)b&	1.78&	0.03&	6.05&	4&	141.1&	1.58E-012&	1.06&	0.009&	WFPC2\_F814\\
NGC1386&	10.6&	S2&	SB0+(s) &	1.10&	0.02&	7.23&	2&	133.1&	8.70E-013&	-0.11&	0.001&	WFPC2\_F814\\
NGC1399&	19.4&	S2&	E1 pec&	0.58&	0.01&	8.94&	3&	332.2& $-$ & $-$ & $-$ &			WFPC2\_F814\\
NGC1433&	13.3&	S2&	(R')SB(r)ab&	0.91&	0.02&	7.24&	5&	107&	6.07E-014&	0.38&	0.002&	WFPC2\_F814\\
NGC1566&	19.4&	S1.5&	SAB(s)bc&	1.03&	0.03&	7.11&	4&	97.7&	8.88E-014&	0.67&	0.005&	WFPC2\_F814\\
NGC1667&	61.2&	S2&	SAB(r)c&	1.09&	0.03&	7.88&	2&	169.4&	9.28E-014&	1.51&	0.049&	WFPC2\_F606\\
NGC1672&	16.7&	S2&	SB(s)b&	0.77&	0.02&	7.70&	6&	109.5&	$-$ &	1.35&	0.010&	WFPC2\_F814\\
NGC1808&	12.3&	S2&	(R)SAB(s)a&	0.90&	0.02&	7.20&	7&	126.1& $-$ &		-0.11&	0.001&	WFPC2\_F814\\
NGC2273&	28.4&	S2&	SB(r)a&	1.15&	0.03&	7.30&	2&	141&	1.47E-013&	0.14&	0.004&	WFPC2\_F791\\
NGC2639&	42.6&	S1.9&	(R)SA(r)a&	0.69&	0.03&	7.94&	2&	175.3&	3.27E-014&	0.43&	0.009&	WFPC2\_F814\\
NGC2782&	37&	S2&	SAB(rs)a pec&	1.21&	0.03&	7.70&	8&	182.2&	$-$ &	1.81&	0.032&	WFPC2\_F814\\
NGC2974&	20.9&	S2&	E4&	1.20&	0.02&	8.23&	9&	232.2&	$-$ & $-$ & $-$ &			WFPC2\_F814\\
NGC3081&	34.2&	S2&	(R)SAB(r)0/a&	0.88&	0.03&	7.20&	3&	118.8&	9.89E-013&	$-$ & $-$ &		WFPC2\_F814\\
NGC3169&	17.4&	S&	SA(s)a pec&	0.74&	0.02&	8.01&	10&	184.9&	$-$ &	-0.42&	0.001&	WFPC2\_F814\\
NGC3185&	21.3&	S2&	(R)SB(r)a&	1.58&	0.03&	6.52&	2&	76.1&	4.70E-014&	-0.44&	0.001&	WFPC2\_F814\\
NGC3227&	20.6&	S1.5&	SAB(s)a pec&	1.66&	0.00&	6.75&	11&	126.8&	5.71E-013&	$-$ & $-$ &		ACS\_F814\\
NGC3281&	44.7&	S2&	SA(s)ab pec&	0.86&	0.04&	7.28&	2&	168.6&	1.39E-012& $-$ & $-$ &			WFPC2\_F606\\
NGC3486&	7.4&	S2&	SAB(r)c&	1.10&	0.03&	7.00&	12&	60.2&	3.30E-014&	-0.41&	0.001&	WFPC2\_F814\\
NGC3489&	6.73&	S2&	SAB0+(rs)&	1.34&	0.01&	6.78&	13&	104.2&	$-$ &	-0.41&	0.001&	WFPC2\_F814\\
NGC3516&	38.9&	S1.2&	(R)SB(s)0&	1.45&	0.02&	7.23&	24&	153.6&	5.60E-013&	1.83&	0.037&	WFPC2\_F791\\
NGC3982&	17&	S1.9&	SAB(r)b&	1.12&	0.03&	7.20&	14&	71.8&	1.18E-013&	$-$ & $-$ &		WFPC2\_F814\\
NGC4168&	16.8&	S1.9&	E2&	0.74&	0.04&	9.01&	15&	182&	1.39E-014&	$-$ & $-$ &		WFPC2\_F702\\
NGC4235&	35.1&	S1.2&	SA(s)a&	0.99&	0.03&	7.64&	2&	133.6&	4.33E-014&	-0.17&	0.003&	WFPC2\_F606\\
NGC4258&	8&	S1.9&	SAB(s)bc&	0.72&	0.08&	7.58&	2&	132.8&	7.49E-014&	0.25&	0.001&	ACS\_F814\\
NGC4303&	13.6&	S2&	SAB(rs)bc&	1.69&	0.01&	6.58&	16&	95.1& $-$ & $-$ & $-$ &			WFPC2\_F814\\
NGC4374&	18.5&	S2&	E1&	0.60&	0.01&	8.97&	1&	277.6&	$-$ & $-$ & $-$ &			WFPC2\_F814\\
NGC4472&	17.1&	S2&	E2&	0.53&	0.02&	9.18&	17&	282&	6.64E-014& $-$ & $-$ &			WFPC2\_F814\\
NGC4477&	16.8&	S2&	SB0(s)&	0.79&	0.02&	7.91&	2&	172.5&	1.69E-014&	0.08&	0.002&	WFPC2\_F606\\
NGC4501&	16.8&	S2&	SA(rs)b&	1.22&	0.02&	7.30&	3&	166.2&	3.98E-014&	0.02&	0.004&	WFPC2\_F606\\
NGC4507&	59.6&	S2&	(R')SAB(rs)b&	1.34&	0.02&	7.58&	2&	149&	3.31E-013& $-$ & $-$ &			WFPC2\_F814\\
NGC4552&	15.4&	S2&	E0-1&	0.91&	0.01&	8.63&	18&	250.3& $-$ & $-$ & $-$ &				WFPC2\_F814\\
NGC4579&	16.8&	S1.9&	SAB(rs)b&	1.13&	0.02&	7.70&	19&	165.8&	2.83E-014&	0.22&	0.004&	WFPC2\_F791\\
NGC4594&	20&	S1.9&	SA(s)a&	0.91&	0.01&	8.76&	2&	225.7&	2.62E-014&	0.38&	0.003&	WFPC2\_F814\\
NGC4698&	16.8&	S2&	SA(s)ab&	1.03&	0.02&	7.61&	2&	137.4&	2.03E-014&	-0.53&	0.001&	WFPC2\_F814\\
NGC4725&	12.4&	S2&	SAB(r)ab pec&	0.87&	0.01&	7.51&	2&	131.5&	1.24E-014&	-0.01&	0.002&	WFPC2\_F606\\
NGC5005&	21.3&	S2&	SAB(rs)bc&	0.60&	0.00&	7.84&	20&	171.5&	1.99E-014&	1.91&	0.013&	ACS\_F814\\
\hline		
\end{tabular}
\end{center}

\end{table}

\begin{table}
\addtolength{\tabcolsep}{-3pt}
\begin{center}
\label{tab:continued} 


\begin{tabular}{lcccccccccccc} 
\hline
\vspace{-1mm}
Galaxy & Distance & Seyfert & Morphology & CIR & $\Delta$CIR & log $M_{\rm BH}$ & ref & $\sigma$  & OIV flux  & log SFR & $\Delta$SFR & HST  \\
 &(Mpc) & type &  & &  & $( M_{\odot})$ &  &  $\rm{(km\,s^{-1})}$ &$\rm{(erg\,cm^{-2}\,s^{-1})}$& $(M_{\odot}\,yr^{-1})$ & & obs.  \\
\hline

NGC5033&	18.7&	S1.9&	SA(s)c&	0.96&	0.02&	7.64&	2&	133.9&	1.59E-013&	-0.28&	0.001&	WFPC2\_F814\\
NGC5135&	57.7&	S2&	SB(s)ab&	1.11&	0.02&	7.34&	2&	125.5&	5.83E-013&	1.95&	0.069&	WFPC2\_F606\\
NGC5194&	8.4&	S2&	SA(s)bc pec&	1.27&	0.02&	6.95&	2&	87.9&	2.46E-013&	-0.62&	0.000&	WFPC2\_F814\\
NGC5273&	21.3&	S1.5&	SA0(s)&	1.47&	0.03&	6.61&	3&	65.9&	3.72E-014&	-0.10&	0.001&	WFPC2\_F791\\
NGC5322&	24.3&	S&	E3-4&	0.96&	0.02&	8.51&	21&	230&	$-$ & $-$ & $-$ &			WFPC2\_F814\\
NGC5427&	40.4&	S2&	SA(s)c pec&	1.16&	0.03&	7.58&	3&	69.9&	2.68E-014&	1.00&	0.013&	WFPC2\_F606\\
NGC5643&	14.4&	S2&	SAB(rs)c&	1.45&	0.02&	6.44&	22&	$-$ &	8.16E-013&	$-$ & $-$ &		WFPC2\_F814\\
NGC5806&	27.4&	S2&	SAB(s)b&	1.10&	0.03&	7.07&	23&	124.3&	$-$ & $-$ & $-$ &			WFPC2\_F814\\
NGC6814&	25.6&	S1.5&	SAB(rs)bc&	1.29&	0.03&	7.26&	2&	108.1&	2.13E-013&	1.55&	0.025&	WFPC2\_F606\\
NGC6951&	24.1&	S2&	SAB(rs)bc&	0.73&	0.02&	7.34&	2&	114.8&	8.37E-014&	0.60&	0.014&	WFPC2\_F814\\
NGC7469&	67&	S1.2&	(R')SAB(rs)a&	1.06&	0.00&	7.08&	2&	132.9&	3.67E-013&	2.99&	0.281&	ACS\_F814\\
NGC7479&	32.4&	S1.9&	SB(s)c&	0.85&	0.04&	7.68&	2&	151.3&	2.67E-013&	$-$ & $-$ &		WFPC2\_F814\\
NGC7582&	22&	S2&	(R')SB(s)ab&	1.84&	0.02&	7.74&	2&	118.4&	2.22E-012&	-0.10&	0.002&	WFPC2\_F606\\
NGC7590&	22&	S2&	SA(rs)bc&	1.09&	0.02&	6.79&	2&	91.5&	6.88E-014&	0.52&	0.005&	WFPC2\_F606\\
NGC7743&	24.4&	S2&	(R)SB0+(s)&	1.49&	0.02&	6.72&	2&	83.5&	3.30E-014&	$-$ & $-$ &		WFPC2\_F606\\
\hline		
\end{tabular}

\begin{flushleft}
	\vspace{-2mm}
	\footnotesize
References:(1)\cite{2013ARA&A..51..511K} (2)\cite{2012ApJ...746..168D} (3)\cite{2016ApJ...831..134V} (4)\cite{2014ApJ...789..124D} (5)\cite{2014A&A...567A.119S} (6)\cite{2019A&A...623A..79C} (7)\cite{2017A&A...598A..55B} (8)\cite{2006AJ....131.1236D} (9)\cite{2016ApJS..222...10S} (10)\cite{2015MNRAS.453.3447D} (11)\cite{2003ApJ...585..121O} (12)\cite{2017A&A...601A..20K} (13)\cite{2010MNRAS.403..646N} (14)\cite{2012MNRAS.419.2497B} (15)\cite{1998AJ....115.2285M} (16)\cite{2018ApJ...869..113D} (17)\cite{2019MNRAS.484..794G} (18)\cite{2010ApJ...717..640P} (19)\cite{2001ApJ...546..205B} (20)\cite{2016ApJ...827...81I} (21)\cite{2018MNRAS.475.4670D} (22)\cite{2010MNRAS.406..597G} (23)\cite{2007MNRAS.379.1249D}.
\end{flushleft}
\end{center}

\end{table}

\section{The Data and Data Reduction}
\label{sect:Obs}

For this work, we consider a complete sample of  Seyfert galaxies drawn from the Revised Shapley-Ames catalog \citep[RSA][]{1932AnHar..88...41S,1987rsac.book.....S} analyzed by \cite{2009ApJ...698..623D,2012ApJ...746..168D}, which include 114 nearby ($z<0.02$) Seyfert galaxies brighter than $ B_T =$ 13.2. We took archival images of the sample observed by the Hubble Space Telescope (HST) to estimate CIR. Among the 114 galaxies, 83 had HST observations in optical bands. From this, 13 galaxies consisting of bad pixels or defects on their images within the region of central 3 arcsecs are avoided from the sample. Also, we excluded 10 small galaxies, in which the size of their images is less than the 3 arcsecs aperture, and 3 highly inclined galaxies ($i>70^{\circ}$). The final sample consists of 57 Seyfert galaxies comprising 40 spiral, 9 lenticular, and 8 elliptical galaxies. We chose galaxy images observed with Wide Field Planetary Camera 2 (WFPC2). It is already reported that though there exist slight variations among values of CIR estimated from nearby filters, the variations are not strong enough to disrupt the observed correlations involving CIR \citep{2021MNRAS.500.1343S}. Hence we included observations using all filters of WFPC2 from F606W to F814W, giving preference to the filter in the highest wavelength region available. Further, we added four observations using F814W with Advanced Camera for Surveys to improve the statistics for which no  WFPC2 images were available. Since the sample galaxies possess AGN, we have checked the central region of the sample galaxies by constructing residual images using the \emph{ellipse} task in IRAF. Many sub structures like bar, ring, spiral-arms are visible in the residual image, however, no significant optical excess flux could be identified in these galaxies that could affect determination of CIR.

Following  \cite{2018MNRAS.477.2399A}, the CIR  is determined for the sample galaxies by using the aperture photometry ($\rm{MAG\_APER}$) technique, which is provided in source extractor \citep[SExtractor,][]{1996A&AS..117..393B}.

\begin{equation}
{\rm CIR} = \frac{I_1}{I_2-I_1}=\frac{10^{0.4(m_2-m_1)}}{1-10^{0.4(m_2-m_1)}}
\end{equation}

where $ I_1$ and $ I_2$ are the intensities and $ m_1$ and $ m_2$ are the corresponding magnitudes of the light within the inner  and outer  apertures of radii $ R_1$ and $ R_2$, respectively. The inner radius is chosen such that its a few times the PSF. The outer radius (conventionally $2 R_1$) is chosen such that the aperture is lying fairly within the galaxy image.  For the sample, we chose the inner and outer radii as 1.5 and 3 arcsecs, respectively.

Ultraviolet (UV) observations are vital in providing recent star formation activity in galaxies \citep[e.g.,][]{2005ApJ...627L..29G,2007ApJ...661..115G,2005ApJ...619L..79T,2009MNRAS.400.1749K}.  For the estimation of circumnuclear star formation rate (SFR), we took far-UV (FUV, 1350-1750 \AA) data of the sample galaxies observed by the GALaxy Evolution eXplorer (GALEX) mission.  We used an aperture size of 10 arcsecs at the centre of the image to estimate the circumnuclear SFR following \cite{2010A&A...521A..63L} using the calibration provided by \cite {2007ApJS..173..267S} and is provided in Table \ref{Tab1}.

\begin{figure}
\hspace*{-6mm}
\includegraphics[width=0.8\linewidth]{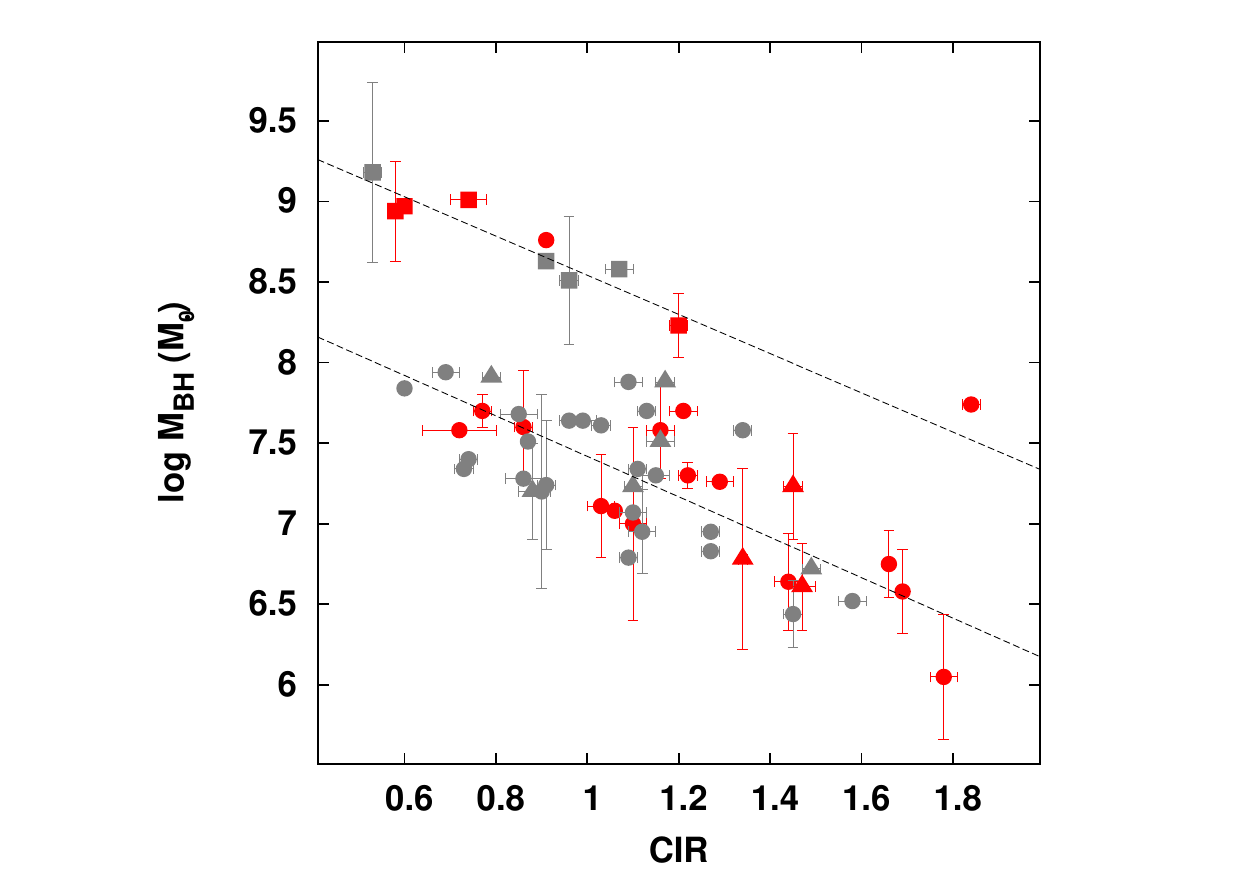}
\caption{Correlation between the central intensity ratio and mass of the SMBH of the sample galaxies. Filled circles, triangles, and squares represent spiral, lenticular, and elliptical galaxies, respectively. Estimations of masses of SMBH using dynamical methods are denoted as red data points, while gray color is used to represent those from the stellar velocity dispersion measurements of host galaxies.}
\label{fig:1}
\end{figure}

\section{Results}

Scaling relations displayed by various structural and dynamical observables of galaxies can shed vital information on formation and evolution processes in galaxies. We have estimated the CIR at the optical centre of 57 Seyfert galaxies observed using HST. The sample properties, along with the estimated values of CIR, are tabulated in Table \ref{Tab1}. We next explore various trends involving CIR. 

\subsection{Variation of CIR with SMBH mass}
The scaling relations of black hole mass are generally determined and explored utilizing the bulge properties of the host galaxies, specifically in early-type galaxies \citep{1995ARA&A..33..581K,2000ApJ...539L...9F,2013ApJ...764..184M}. Structural properties of late-type galaxies, like the pitch angle of spiral arms, share intriguing scaling relation with the black hole mass \citep{2018ApJ...869..113D,2019ApJ...873...85D}.

Figure \ref{fig:1} shows the variation of CIR with mass of SMBH for the sample galaxies. Filled circles, triangles, and squares represent spiral, lenticular, and elliptical galaxies, respectively.  We find a strong correlation between the CIR of the Seyfert galaxies and the mass of their central SMBH. However, the early-type galaxies in the sample host systematically high black hole mass when compared to lenticulars and spirals, for the same value of CIR considered.  The Pearson's linear correlation coefficient, $r$, for the correlation exhibited by spirals and lenticulars together is  -0.74 with a significance, $s$, $> 99.99$ per cent \citep{1992nrfa.book.....P} while that for elliptical galaxies is -0.94 with a significance of 99.40 per cent.

Two galaxies, NGC4594 and  NGC7582, show significant deviation from this correlation. NGC4594, the Sombrero galaxy, is reported to have an unusually large bulge mass and a very massive SMBH at the centre of the galaxy. It is usually classified as a normal spiral, Sa, galaxy \citep{1991rc3..book.....D} but following many scaling relations of ellipticals \citep{Gadotti2012}.
NGC7582 is reported to host a ring with active star formation within the pc scale radius ($\approx 190$~pc) surrounding the nucleus of the galaxy, along with a high stellar velocity dispersion \citep{2009MNRAS.393..783R}. The intense nuclear starburst activity \citep{2001RMxAC..11..133C,2007MNRAS.374..697B} can affect its CIR value.

In Section \ref{secn-cir-sigma}, we notice that there is no apparent correlation between CIR and stellar velocity dispersion of host galaxies in our sample, even though the latter and mass of SMBH are reported to share a strong correlation.  In order to explore this discrepancy, we also employed a color code to distinguish the method adopted to estimate the masses of SMBH. Masses estimated using a dynamical method (e.g., reverberation mapping, stellar dynamics, maser dynamics and gas dynamics) are shown in red while mass estimations based on stellar velocity dispersion are shown in gray in Figure \ref{fig:1}.  If we include only dynamically estimated masses, the correlation coefficient improves to -0.77 at a significance, $s = $ 99.97 per cent, while it reduces drastically to -0.68 ($s = $ 99.98 per cent) when these data points are excluded.
 
\subsection{Variation between the CIR and $\sigma$}
\label{secn-cir-sigma}
The variation of  CIR with the stellar velocity dispersion of the sample galaxies is shown in Figure \ref{fig:2}(a). As already mentioned, there is no significant correlation between CIR and stellar velocity dispersion ($\sigma$) of Seyfert galaxies. However, if we exclude the eight early-type galaxies in the sample, the velocity dispersion measurements of galaxies with dynamical estimation of SMBH (red triangles and circles) show larger scatter compared to their gray counterparts. Such a discrepancy is not clear in early-type galaxies. The extreme emission from AGN activity can complicate the measurement of central velocity dispersion in these galaxies  \citep{2013MNRAS.429.2587R}.

Measurements of stellar velocity dispersion may be biased by the contribution of rotating stellar disks because of the rotational broadening of the stellar absorption lines and the velocity dispersion measurements could be noticeably increased by the rotational effect \citep{2015ApJ...801...38W}. Due to higher velocity-to-dispersion (V/$\sigma$) ratios, the rotational effect is significantly more prominent in late-type galaxies (LTGs) than in ETGs. 

\subsection{Variation between the CIR and SFR}
In Figure \ref{fig:2}(b), we explore the connection between CIR and circumnuclear SFR traced by the UV luminosity (FUV) in an aperture of radius 10 arcsec at the galactic centre. We find that there is no correlation between CIR and circumnuclear SFR. The properties of sub-structures in the nuclear region of host galaxies may influence the star formation process, and there by affecting CIR.  The galaxies IC2560, NGC0788, NGC1667, NGC3516, NGC5427, and NGC6814, denoted by the numbers 1 to 6 respectively in the figure, possess nuclear dust spirals, which can regulate the nuclear SFR at the central region of the galaxies  \citep{2007AJ....134..648M,1996ApJS..105...93E,2003ApJ...589..774M,2000MNRAS.317..234P}. The galaxies IC3639, NGC2782, NGC5135, and NGC7582, numbered 7 to 10, with nuclear start burst activity \citep{1992A&A...260...67B,2001ApJ...546..845G,2007AJ....134..648M,2007MNRAS.374..697B} are also apparent outliers in the figure. NGC1365 and NGC7469 are the galaxies showing intense nuclear SFR, with star-forming regions concentrated in hot spots around the nucleus \citep{2009ApJ...702.1127R,2007ApJ...671.1388D}, which are shown in the figure by the numbers 11 and 12. By excluding these galaxies, we may see a negative trend in CIR and SFR. However, it is insufficient to confirm any connection between CIR and SFR, necessitating a thorough investigation with larger sample size.

\subsection{Variation between the CIR and OIV flux}
\begin{figure*}
    \centering
    \begin{subfigure}[h!]{0.45\linewidth}
        \includegraphics[height=70mm,width=90mm,angle=0]{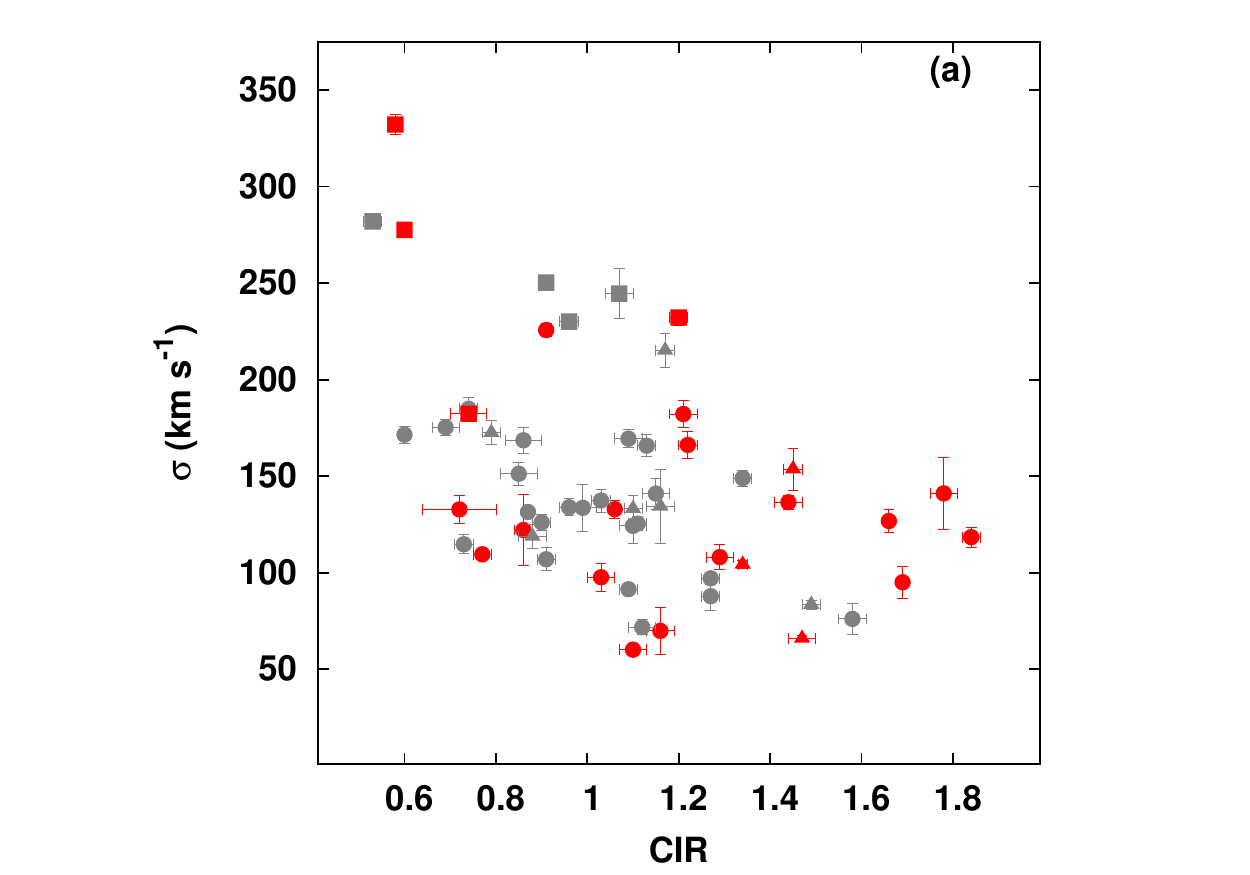}
    \end{subfigure}
        \vspace{4mm}
        \hspace*{3mm}
    \begin{subfigure}[h!]{0.45\linewidth}
        \includegraphics[height=70mm,width=90mm,angle=0]{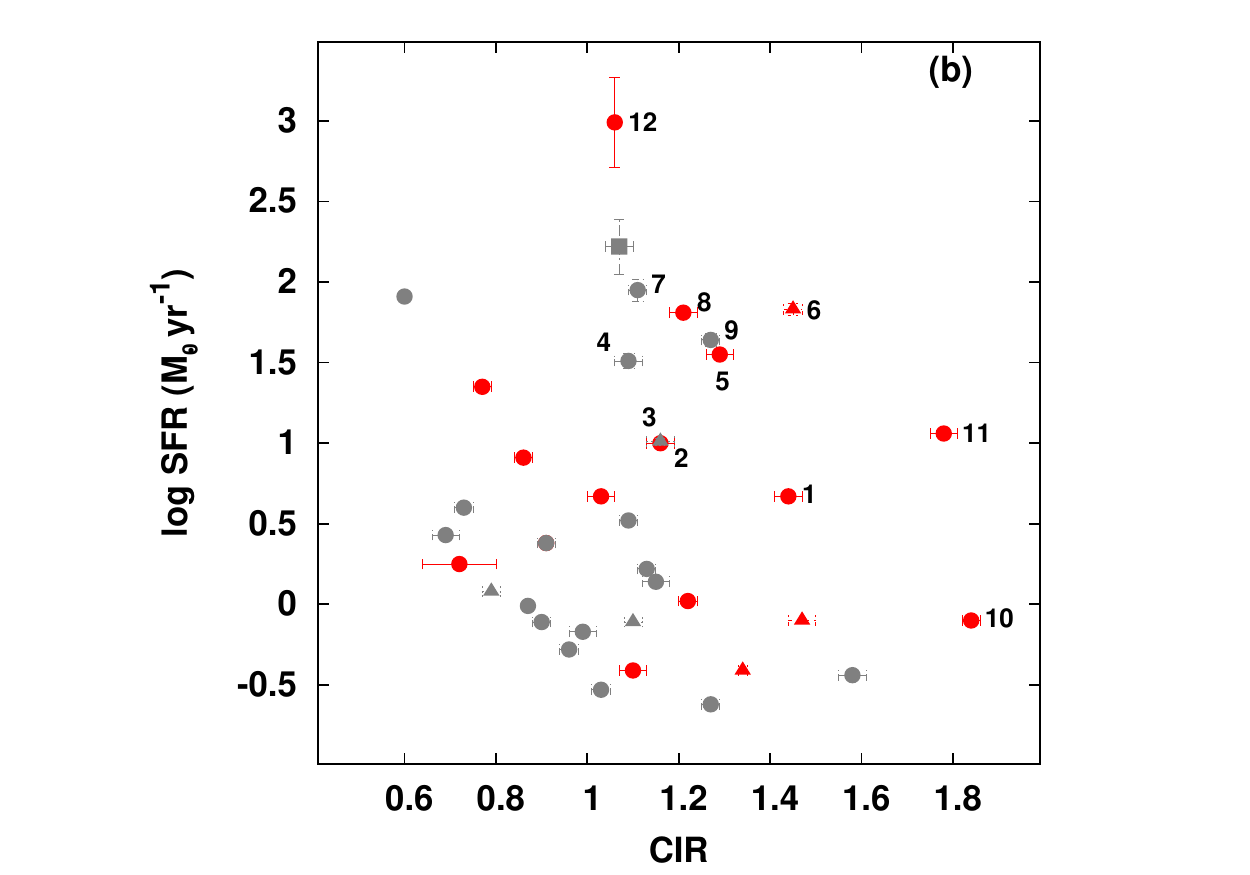}
    \end{subfigure}
    \vspace{4mm}
    \hspace*{3mm}
    \centering
    \hspace*{2mm}
    \begin{subfigure}[h!]{0.45\linewidth}
        \includegraphics[height=70mm,width=90mm,angle=0]{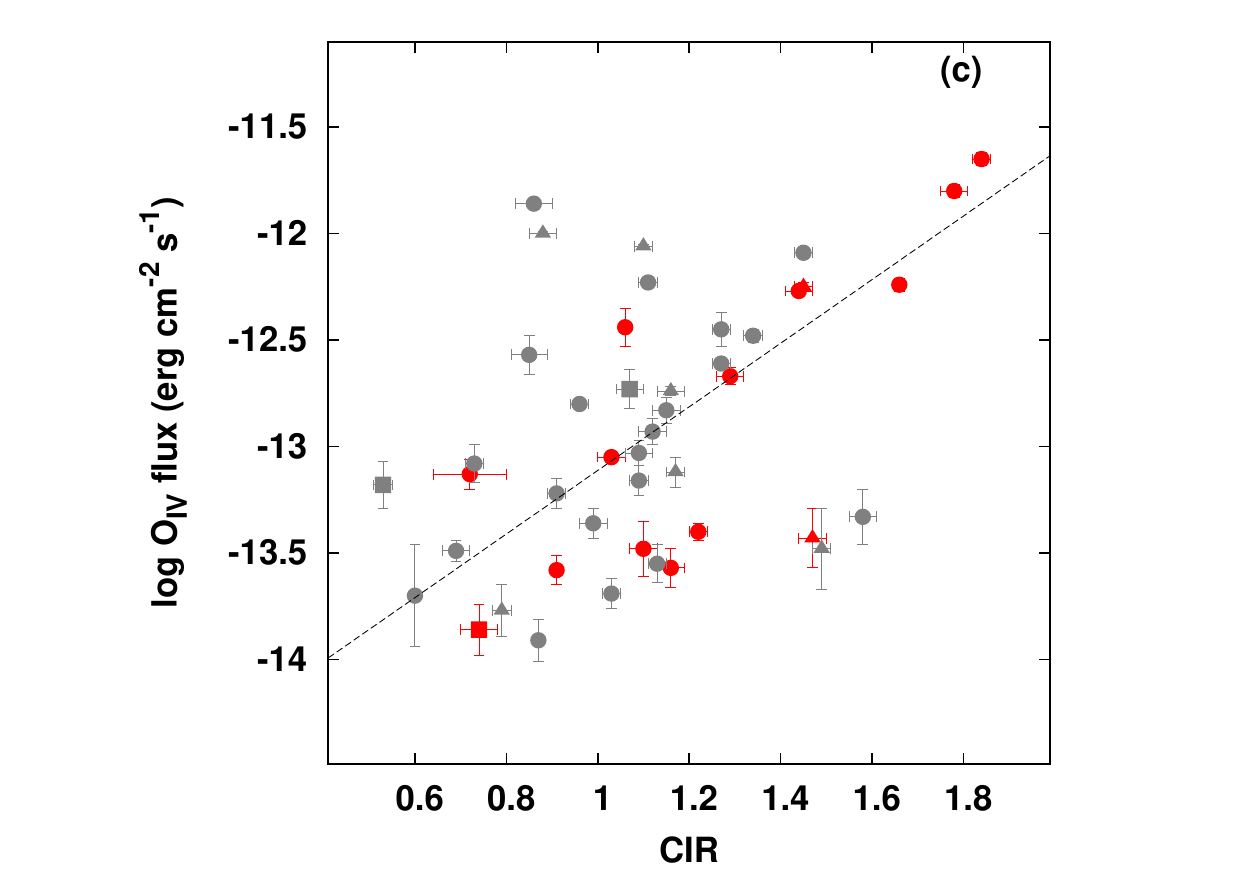}
    \end{subfigure}
        \vspace{4mm}
        \hspace*{3mm}
    \begin{subfigure}[h!]{0.45\linewidth}
        \includegraphics[height=70mm,width=90mm,angle=0]{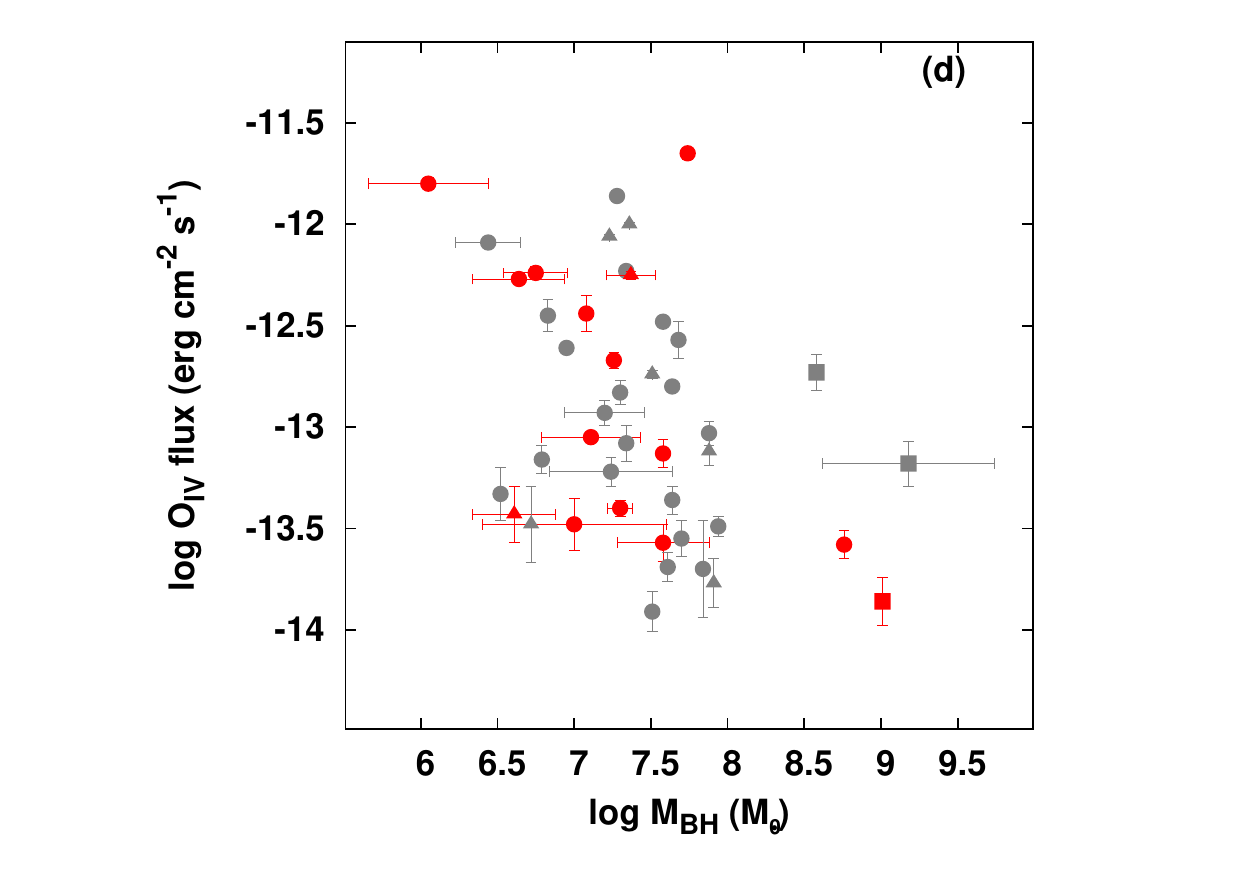}
    \end{subfigure}
    \vspace{4mm}
    \hspace*{3mm}
        \caption{Variations between the central intensity ratio and (a)stellar velocity dispersion adopted from HyperLEDA database (b)circumnuclear star formation rate (c)OIV flux of AGN and (d)the inter-connection between 
        $ M_{\rm BH}$ and OIV flux of the sample galaxies. OIV flux values are taken from \citep{2009ApJ...698..623D}. The symbols used to denote the galaxies are same as Fig. \ref{fig:1}.}
        \label{fig:2}
\end{figure*}

In Figure \ref{fig:2}(c), we show the observed correlation between CIR and OIV flux of the host galaxy, which is taken from \cite{2009ApJ...698..623D}. OIV flux is an accurate measure of intrinsic AGN luminosity \citep{2009ApJ...698..623D} and we find a positive correlation with CIR ($r = 0.70$ with $s >$ 99.99 per cent). OIV emission (25.9~${\rm \mu}$m) is a tracer of highly ionized gas of the order of 35-97 eV, and these types of mid-IR emission lines can be produced in the vicinity of hot stars in the central region of AGN host galaxies \citep{2001A&A...380..684P,2004ApJS..154..199S,2007AAS...21011209D}. AGN luminosity depends upon the fuel consumed by the SMBH at the nuclear region of the galaxy, and early-type galaxies have less fuel than the late-type galaxies \citep{2002ASPC..258..113R}. This suggests that AGN power is likely to decrease while SMBH grows in the host galaxy. However, NGC3081, NGC3185, NGC3281, NGC5273, and NGC7743 deviated from this correlation.

%

\begin{figure}
\hspace*{-6mm}
\includegraphics[width=0.8\linewidth]{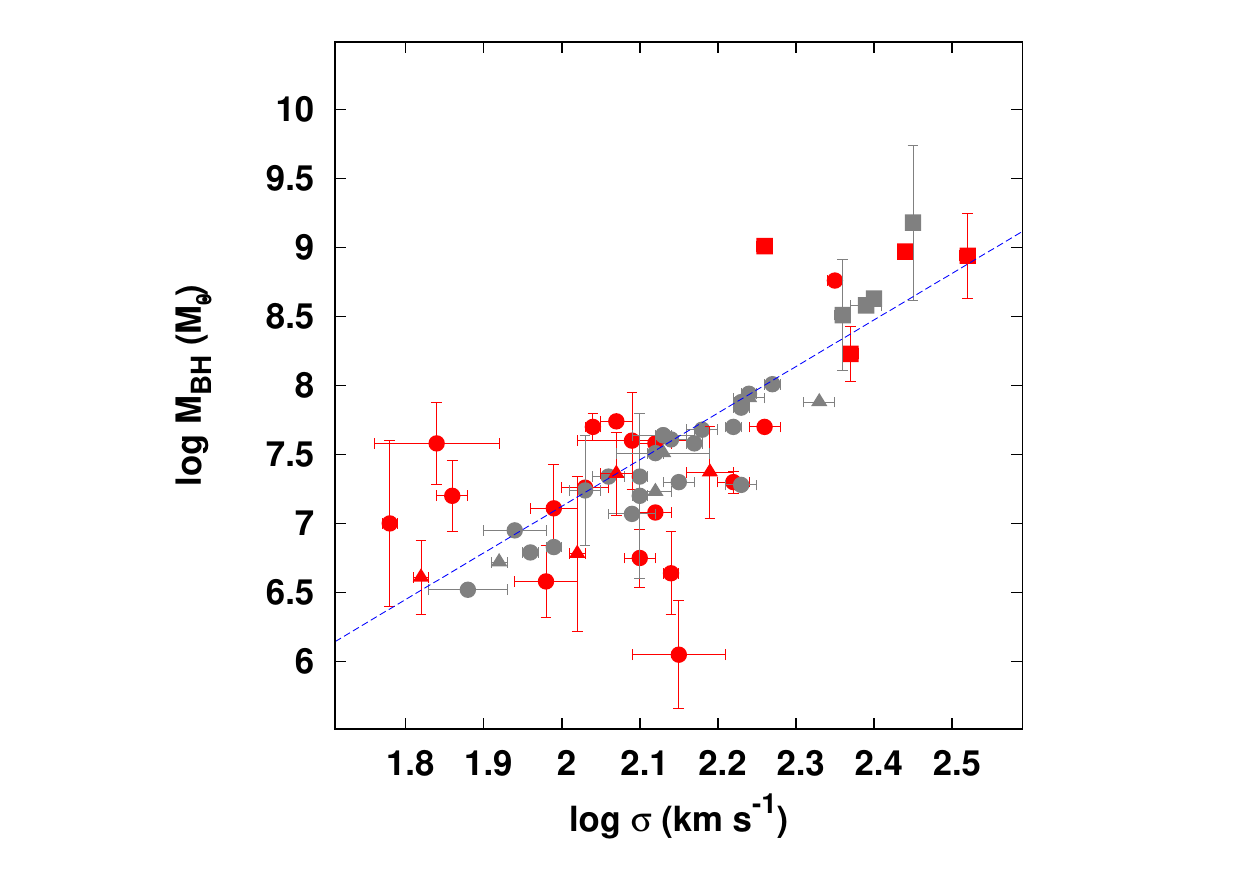}
\caption{Variation between stellar velocity dispersion and mass of the SMBH of sample galaxies along with the best fit adopted from \cite{2020A&A...634A.114C}. The symbols used to denote the galaxies are same as in Fig. \ref{fig:1}.}
\label{fig:3}
\end{figure}

\begin{table}
\addtolength{\tabcolsep}{0pt}
\begin{center}
\caption{The table lists the best-fitting parameters for the relation $\rm{x = \alpha~CIR + \beta}$ and correlation coefficients for various relations.}          
\begin{tabular}{l c c c c c }
\hline
 x	&	$\alpha$	& $\beta$	&	$r$	&	$s$	&	n \\
\hline
log $ M_{\rm BH}$ &  -1.25$\pm$0.16 & 8.67$\pm$0.19 & -0.74 & $ >99.99\%$ & 47\\
(spiral+lenticular)\\
log $ M_{\rm BH}$ &  -1.16$\pm$0.15 & 9.67$\pm$0.13 & -0.94 & $ 99.40\%$ & 8\\
(elliptical)\\
log OIV flux &  1.49$\pm$0.24 & -14.60$\pm$0.28 & 0.70 & $ >99.99\%$ & 38\\
\hline
\end{tabular}
\end{center}
\end{table}

\section{Discussion and Conclusion}
We report photometric analysis of Seyfert galaxies using the recently discovered parameter central intensity ratio. The CIR shows good correlations  with many structural parameters of host galaxies, especially with the mass of the SMBHs residing at the centre of galaxies \citep{2018MNRAS.477.2399A,2020RAA....20...15A}. For Seyfert galaxies also, the CIR shows strong anti-correlation with the mass of SMBHs. However, the massive SMBHs hosted by ellipticals in the sample display a distinct trend from that displayed by lenticulars and spirals, in the sense that ellipticals host more massive SMBHs than that hosted by lenticulars and spirals.  The disky systems are indistinguishable in the correlation. It is possible that the more massive the central SMBH, the higher the suppression of star formation due to feedback \citep{2017NatAs...1E.165H}. As a decrease in the light in the inner aperture reduces the value of CIR, we can expect the anti-correlation between CIR and mass of SMBH.

The AGN feedback mechanism has a significant role in the evolution process of galaxies, in which the energy released by AGN to the surrounding galactic medium  halts the cooling of gas in the central region of galaxies and also removes the gas in the form of massive outflows \citep{2017FrASS...4...42M}. The AGN feedback process is considered a key factor of galaxy evolution and has been included in several simulations and analytical models for years \citep[e.g.,][]{2000MNRAS.311..576K,2005Natur.433..604D,2015MNRAS.446..521S,2015MNRAS.452..575S}. This feedback may suppress the star formation at the central part of the galaxy and may decrease or stall completely the growth of the SMBH \citep[e.g.,][]{2006MNRAS.365...11C,2007MNRAS.380..877S}, thus setting up a co-evolution scenario for the galaxy and its SMBH \citep{2018MNRAS.477.2399A}. Around 30 per cent of Seyfert galaxies are reported to possess outflow incidents  \citep{2007ASPC..373..319C,2003ARA&A..41..117C,2012ASPC..460..261C}. The pc scale AGN-driven outflows in the massive galaxies can expel the gas from the nuclear region, which may reduce the gas accretion towards the centre of the galaxy and leads to quenching of star formation at the central region  \citep{2017FrASS...4...42M}. This interesting phenomenon has been observed in optical, UV, and X-ray emissions and could be traced to such outflows using ionized gas and absorption lines \citep[e.g.,][]{2005ARA&A..43..769V,2007Ap&SS.311...87B,2008MmSAI..79.1205T,2015ARA&A..53..115K}. 

Different studies argued for the probability of AGN feedback by thermal process in the vicinity of the SMBH \citep[e.g.,][]{2005Natur.433..604D,2005MNRAS.361..776S,2009ApJ...690..802J}. Simulations of the AGN feedback mechanism suggest that the Compton heating effect can raise the temperature of the the gas at the nuclear region of about 10-35 pc,  to $\sim10^9$~K \citep[e.g.,][]{2014ApJ...789..150G,2015ApJ...812...90M}. This AGN heating may also reduce the star formation in the central region of the galaxy, thus the value of CIR.

Stellar velocity dispersion $(\sigma)$ of the bulge component is strongly connected with the central SMBH \citep{2000ApJ...539L...9F,2000ApJ...539L..13G,2002ApJ...574..740T,2009ApJ...698..198G}. Active galaxies are also obeying the $\sigma - M_{\rm BH}$ relation, but with significant scatter \citep{2020A&A...634A.114C}.  It is also reported that CIR of  early-type galaxies are well correlated with the stellar velocity dispersion \citep{2018MNRAS.477.2399A}. In the present study, however, Seyfert galaxies show a large scatter in the $\rm CIR - \sigma$ relation, even though there is a strong $\rm CIR$ - $M_{\rm BH}$ relation. The uncertainties present in the  measurement of stellar velocity dispersion could be high when excessively illuminated by central AGN \citep{2013MNRAS.429.2587R}. Furthermore, stellar velocity dispersion measurements may be skewed due to the rotational effect of stellar disks  \citep{2015ApJ...801...38W}.  In order to explain further this, we have plotted the variation of $ M_{\rm BH}$ with  $\sigma$ in Figure \ref{fig:3}. The velocity dispersion measurements for galaxies with dynamical estimation of mass of SMBH available, shown in red, clearly display a larger scatter than that of galaxies without dynamical estimation of $ M_{\rm BH}$ (gray points). For the $\sigma - M_{\rm BH}$  correlation in the combined sample of spirals and lenticulars, we obtained a linear correlation coefficient of  $r = 0.65$ with significance, $s > 99.99$ per cent. At the same time, the correlation coefficient of $\rm CIR$ - $M_{\rm BH}$ (for spirals + lenticulars) relation is, $r = -0.74$ with significance, $s > 99.99$ per cent. The scatter of the correlations $\sigma - M_{\rm BH}$ and $\rm CIR$ - $M_{\rm BH}$ are 2.63 and 2.04 dex respectively, further establishing that the CIR is, in fact, a better tracer of the $ M_{\rm BH}$ than the central velocity dispersion. 

The observed correlation between CIR and OIV flux, shown in Figure \ref{fig:2}(c), also displays the possibility of larger uncertainties present in measurement of emission lines in galaxies associated with AGN \citep[e.g.,][]{2003A&A...409..867L,2006ApJ...640..204A,2007ApJ...656..148A,2009ApJS..182..628V,2009ApJ...698..623D}. In this case also, the correlation coefficient increases to 0.76 with a significance, $s = $ 99.38 per cent if we just consider galaxies with dynamically estimated SMBH masses, but it drops to 0.62 ($s = $ 99.74 per cent) when these data points are excluded. Seyfert galaxies with high OIV flux emission possess enhanced nuclear star formation \citep{2012ApJ...746..168D}, and an increase in CIR is expected in galaxies with increased OIV emission. However, the OIV flux of the sample galaxies shows only a weak anti-correlation ($r = -0.58$ with $s =$ 99.95 per cent) with the mass of SMBH shown in Figure \ref{fig:2}(d), possibly due to the increased uncertainties involved in both the quantities. 

Generally, Seyfert galaxies can be observed and located through the UV emission coming out from the sources \citep{2002ASPC..258..113R}. Apart from age and morphological classification, the common feature of Seyfert galaxies is their intense star formation \citep{2004MNRAS.355..273C,2007MNRAS.380..949S,2007ApJ...671.1388D,2009MNRAS.397..135K}. We explore the variation of the estimated circumnuclear SFR by the excess UV with CIR, as shown in Figure \ref{fig:2}(b). We notice that the galaxies harbouring central structures such as pc scale nuclear dust spiral, nuclear starburst, and the galaxies possessing high SFR are showing large deviation in the observed $\rm CIR-SFR$ relation. The measure of nuclear SFR has been shown to increase from the central region to the outskirts of galaxies \citep{2012ApJ...746..168D,2014ApJ...780...86E}. The outflow from the central part of the galaxy due to the AGN feedback mechanism can interact with the interstellar medium (ISM) effectively \citep{2010ApJ...722..642O,2017MNRAS.465.3291W,2018ApJ...857..121Y}. The feedback-driven outflow of gas enhances the star formation at larger radii from the core of the galaxy \citep{2013MNRAS.431.2350I,2014MNRAS.441.1474I}. This outflow of gas can be responsible for enhancing the circumnuclear star formation rate. All these can affect measurements of both star formation rate and CIR, rendering a weak correlation between the two.  

We employed CIR, to explore the presence of central features in Seyfert galaxies and their role in galaxy evolution. The analysis shows that CIR measured for Seyfert galaxies predicts the mass of central SMBHs even better than the estimates obtained by spectroscopic parameters like the central velocity dispersion. Being a photometric tool, this promises a cheap and fast technique to explore large galaxy samples, which has great potential in observations of new generation facilities like the James Webb Space Telescope.

%
%


\begin{acknowledgements}
We sincerely thank the anonymous referee for her/his comments which had improved the quality of the paper significantly. VKT would like to acknowledge the financial support from the Council of Scientific \& Industrial Research (CSIR), Government of India. We acknowledge the use of the NASA/IPAC Extragalactic Database (NED), \url{https://ned.ipac.caltech.edu/} operated by the Jet Propulsion Laboratory, California Institute of Technology, and the Hyperleda database, \url{http://leda.univ-lyon1.fr/}. We acknowledge the use of data publicly available at Mikulski Archive for Space Telescopes (MAST), \url{http://archive.stsci.edu/} observed by NASA/ESA Hubble Space Telescope and Galaxy Evolution Explorer (GALEX) led by the California Institute of Technology \url{http://galex.stsci.edu/}
\end{acknowledgements}



\bibliographystyle{raa}
\bibliography{bibtex}

  


\label{lastpage}

\end{document}